\documentclass[aps,twocolumn,prl,groupedaddress,showpacs]{revtex4-2}
\usepackage{graphicx,float}
\usepackage[normalem]{ulem}
\usepackage{color}
\usepackage[colorlinks=true,linkcolor=red,citecolor=blue]{hyperref}

\usepackage{dcolumn}
\usepackage{bm}
\usepackage{tikz}
\usepackage{braket}
\usepackage{siunitx}
\usepackage{gensymb}

\usepackage{amsmath,amsfonts,amssymb, times,natbib}
\usepackage[standard]{ntheorem}

\UseRawInputEncoding

\begin{document}

\title{
Bright and Uniform Silicon Carbide Nanophotonic Arrays for Scalable Nanoscale Quantum Sensing
}

\author{Chenyang Liao$^{1,3}$$^{\S}$}
\thanks{$^{\S}$These authors contributed equally to this work.}

\author{Chihang Luo$^{1}$$^{\S}$}
\thanks{$^{\S}$These authors contributed equally to this work.}

\author{Xin Chen$^{1}$}

\author{Cunliang Xin$^{1}$}

\author{Hongfang Zuo$^{1}$}

\author{Haoyu Sun$^{1,3}$}

\author{Fazhan Shi$^{1,2,3}$}

\author{Mengqi Wang*$^{1,2}$}
\email{mqw@ustc.edu.cn}

\author{Ya Wang*$^{1,2,3}$}
\email{ywustc@ustc.edu.cn}

\affiliation{$^1$ Laboratory of Spin Magnetic Resonance, School of Physical Sciences, Anhui Province Key Laboratory of Scientific Instrument Development and Application, University of Science and Technology of China, Hefei 230026, China}
\affiliation{$^2$ Hefei National Research Center for Physical Sciences at the Microscale, University of Science and Technology of China, Hefei 230026, China}
\affiliation{$^3$ Hefei National Laboratory, University of Science and Technology of China, Hefei 230088, China}

\begin{abstract}
Silicon carbide (SiC) spin defects are emerging as powerful platforms for nanoscale quantum sensing due to their near-surface stability, near-infrared emission, and CMOS compatibility. However, poor photon collection efficiency and the reliance on randomly distributed defects hinder the realization of practical, large-scale devices. Here, we demonstrate scalable, ultra-bright, and highly uniform SiC quantum nanophotonic arrays. By combining a self-aligned fabrication process with localized oxygen ion implantation, we deterministically position PL5 and PL6 centers at the exact center of 4H-SiC nanopillars. This architecture yields record-high photon count rates of $\sim$1 Mcps (PL5) and $\sim$2 Mcps (PL6), establishing a new benchmark for SiC color centers while maintaining near-identical brightness across the array. Furthermore, we apply these devices to nuclear magnetic resonance (NMR), successfully demonstrating external nanoscale nuclear-spin detection. These results establish our SiC nanopillars as robust building blocks for scanning quantum probes and scalable sensor networks.

\end{abstract}
\maketitle

\section{Introduction}

Solid-state optically active spin defects, including nitrogen-vacancy (NV) centers in diamond \textsuperscript{\cite{balasubramanian2008nanoscale,maze2008nanoscale,schirhagl2014nitrogen,Degen2017RMP,Du2024RMP}} and spin defects in silicon carbide \textsuperscript{\cite{falk2013polytype,koehl2011room,christle2015isolated}}, are promising platforms for quantum communication, networks, and nanoscale sensing \textsuperscript{\cite {awschalom2018quantum,wolfowicz2021quantum,Pompili2021Science,staudacher2013nuclear,mamin2013nanoscale,Shi2015Science,lovchinsky2016nuclear,zhou2025entanglement}}.
A pivotal strategy for advancing these quantum technologies is to develop photonic architectures that enable large-scale quantum information processing systems\textsuperscript{\cite{wang2020integrated}}. Nevertheless, qubit inhomogeneities and complex device requirements have impeded progress. 
One promising solution is to leverage heterogeneous integration of uniform small chips into hybrid photonic circuits. Recent advances have enabled large-scale integration of diamond quantum chips into wafer-scale AlN photonic circuits, enabling on-chip multiplexed quantum repeaters and alleviating the constraints imposed by the lack of wafer-scale, single-crystal diamond \textsuperscript{\cite{wan2020large}}. 
An alternative approach is to directly exploit materials with mature wafer-scale manufacturing, such as silicon carbide (SiC)\textsuperscript{\cite{Lukin2020NatPhoton}}.  However, achieving scalable and precise integration of atomic-scale color centers into photonic structures in such materials, with highly uniform performance, remains a formidable challenge\textsuperscript{\cite{castelletto2020silicon,lukin2020integrated}}.

Because they are insensitive to charge noise and emit in the near-infrared, various optically active spin defects in SiC have emerged as a promising quantum system for building quantum networks\textsuperscript{\cite{Christle2017PRX,koehl2011room,Lukin2020NatPhoton}}. Beyond these applications, the recently studied PL5 and PL6 centers in the 4H-SiC polytype are particularly promising for nanoscale quantum sensing at room temperature. In bulk crystals, these defects exhibit stable photon count rates (around 100 kcps) and provide high-contrast, room-temperature optically detected magnetic resonance (ODMR) signals \textsuperscript{\cite{Li2022NSR,hu2026high,chen2026oxygen}}. 

Additionally, their near-infrared optical pumping and emission are highly valuable because they can substantially reduce phototoxicity, avoid the background fluorescence often encountered in biological tissues, and are highly compatible with optical fibers\textsuperscript{\cite{li2025non,krumrein2024precise}}. 

\begin{figure*}
\centering 
\includegraphics[width=0.9\textwidth]{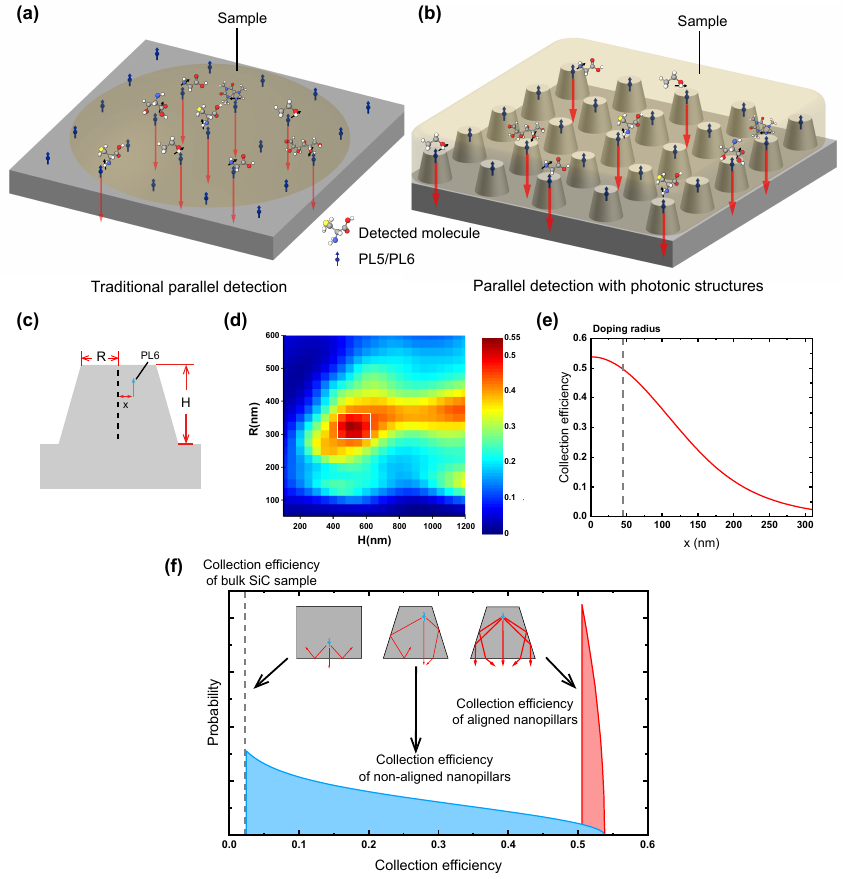}
\caption{ 
(a) Traditional parallel detection of external nuclear spin by PL5/PL6 centers in Silicon Carbide.
(b) Parallel detection of external nuclear spin with photonic structures, using scalable, bright, and uniform SiC nanophotonic arrays.
(c) Simulation model of the nanopillar. The variables R (top radius), H (height), and x (lateral deviation of the color center) are marked in the figure.
(d) The collection efficiency of PL6 as a function of the top radius and the height of the nanopillar, using an air objective with a numerical aperture (NA) of 0.7. The white box marks the optimal parameter range. 
(e) The collection efficiency of PL6 as a function of the lateral deviation of the color center at NA=0.7. The grey dashed line marks the doping radius (45 nm) of the self-aligned nanopillars.
(f) Comparison of collection efficiency among bulk SiC sample, non-aligned SiC nanopillars and self-aligned SiC nanopillars with doping radius of 45 nm. The grey dashed line illustrates the collection efficiency of a 10 nm depth PL6 in the bulk sample. The blue and red lines illustrate the collection efficiency distributions of non-aligned nanopillars and self-aligned nanopillars, respectively, calculated by assuming a uniform distribution of color centers within the doping region and utilizing the results from (e).
}
\label{Figure1}
\end{figure*}

Here, we demonstrate scalable, ultra-bright, and highly uniform SiC nanophotonic arrays capable of nanoscale magnetic detection (Fig.\ref{Figure1}). This is achieved by combining a self-aligned fabrication process\textsuperscript{\cite{wang2022self}} with localized oxygen ion implantation\textsuperscript{\cite{chen2026oxygen}}, which allows shallow PL5 and PL6 centers to be placed precisely at the center of optically designed 4H-SiC nanopillars in a scalable manner. By positioning centers at these optimal sites, we observe photon count rates of $\sim$1 Mcps (PL5) and $\sim$2 Mcps (PL6) with near-identical brightness across the array, setting a new benchmark for photon count rates across all reported SiC color centers \textsuperscript{\cite{widmann2015coherent,radulaski2017scalable,sardi2020scalable,zhou2023plasmonic,wornle2026scalable,chen2026oxygen,koehl2011room}}.

Our results thus offer a route to uniform high-performance optical devices, in contrast to previous work either without nanophotonic structures, which suffers from low photon collection due to the high refractive index (Fig.\ref{Figure1}f, dashed line), or with photonic structures lacking precise alignment, which exhibit random performance owing to the random emitter position (Fig.\ref{Figure1}f, blue line). Applying these devices to nanoscale nuclear magnetic resonance (NMR), we further demonstrate external nuclear-spin detection at the nanoscale under ambient conditions. These results offer a promising route for large-scale nanoscale quantum sensing.

\begin{figure*}
\centering 
\includegraphics[width=0.9\textwidth]{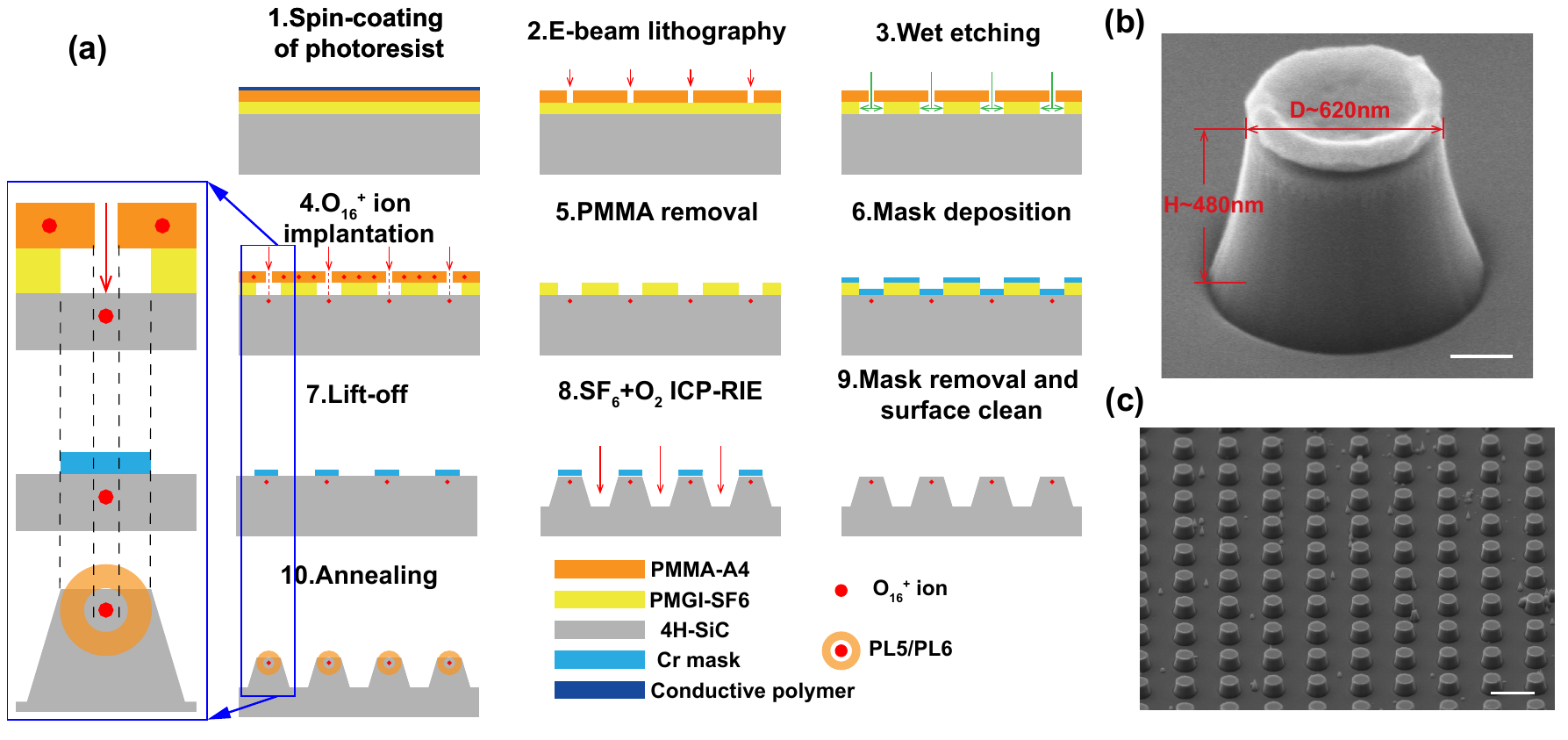}
\caption{
The self-aligned fabrication process for nanophotonic arrays.
(a)Schematic of the fabrication process.
1. Spin-coating of PMGI-SF6, PMMA-A4, and conductive polymer layers on SiC. 2. Electron beam lithography and development to make holes in the PMMA layer. 3. Isotropic wet etching of the PMGI layer. 4.Ion implantation of $\mathrm{O_{16}^+}$ ions through holes in PMMA layer. 5. PMMA layer removal. 6. Deposition of a chromium mask by electron beam evaporation. 7. Lift-off of the mask. 8.$\mathrm{S F_6 +O_2}$ reactive ion etching with inductively coupled plasma (ICP-RIE) to form nanopillars on SiC. 9. Mask removal and surface cleaning in triacid, piranha, and HF. 10. Formation of PL5/PL6 by annealing.
(b)SEM image of a nanopillar before mask removal (scale bar=$200 nm$). The red lines mark the top diameter and height of the nanopillar.
(c)SEM image of a nanopillar nanophotonic array (scale bar=2 $\mu m$).
}
\label{Figure2}
\end{figure*}

\section{Results}

\begin{figure*}
\centering 
\includegraphics[width=0.9\textwidth]{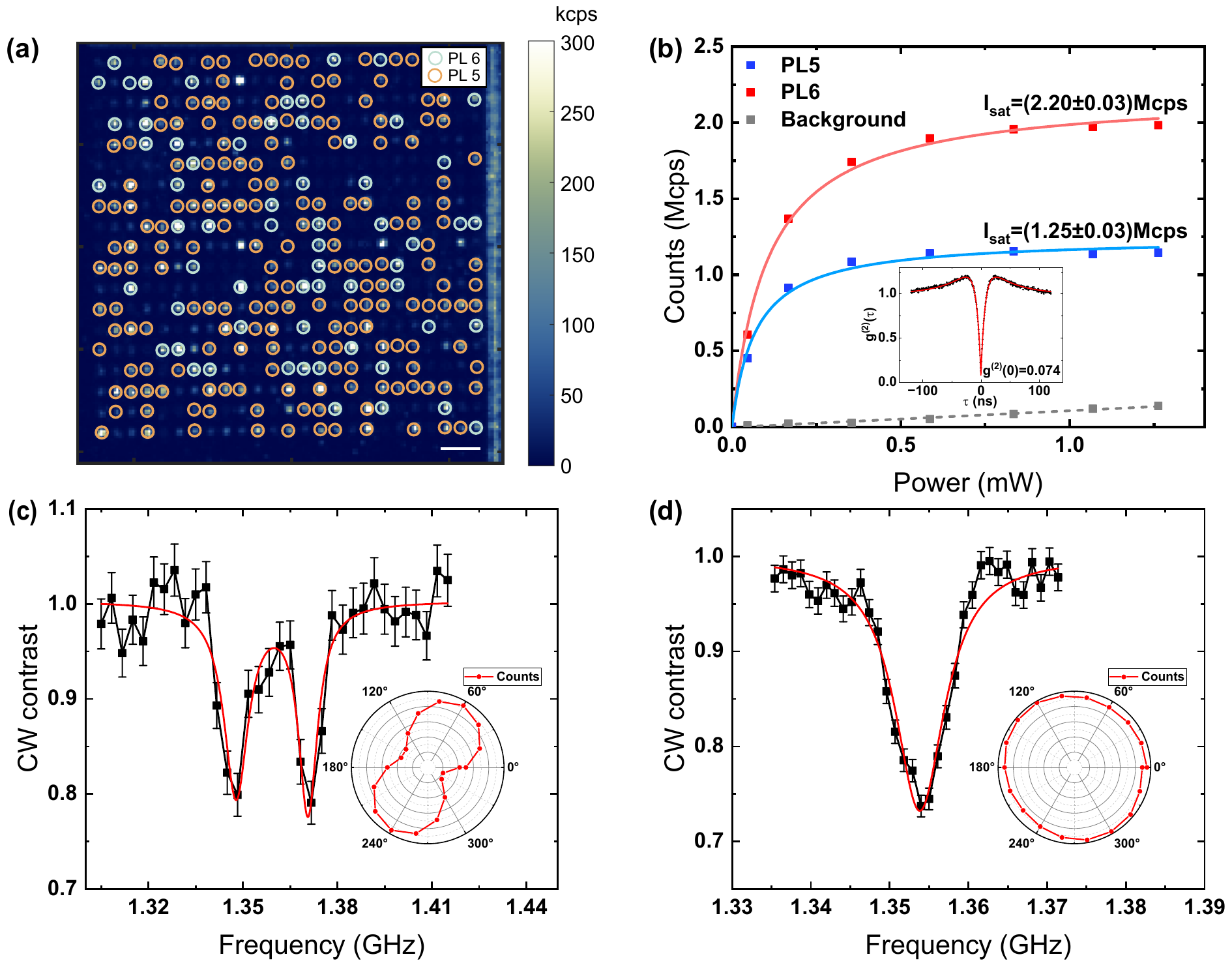}
\caption{Characterization of nanophotonic arrays. 
(a) Confocal image of a nanophotonic array (scale bar=5 $\mu m$). PL5 and PL6 centers are marked in the figure, respectively, by distinguishing their ODMR signals and fluorescence polarization.
(b) The saturation counting rate curves of PL5 (blue) and PL6 (red). The grey dashed line shows the background counting rate. A typical measured second-order correlation function is illustrated in the figure.
(c) ODMR spectra of PL5. Black points are ODMR contrast data, and the red line is a double Lorentzian fit. Error bars represent standard errors of measured data. The dependence of the photon counting rate of PL5 on the excitation light polarization is illustrated in the figure. 
(d) ODMR spectra of PL6. Black points are ODMR contrast data, and the red line is a Lorentzian fit. Error bars represent standard errors of measured data. The dependence of the photon counting rate of PL6 on the excitation light polarization is illustrated in the figure. 
 }
\label{Figure3}
\end{figure*}

\subsection{Nanophotonic Sensor Design and Self-aligned Fabrication}

We fabricate nanopillar optical arrays for enhanced fluorescence collection. Such structures have been applied to NV centers in diamond, where they serve as nanoscale sensing probes for biological magnetic resonance and scanning magnetic imaging \textsuperscript{\cite{babinec2010diamond,wang2019nanoscale,maletinsky2012robust}}. To maximize fluorescence readout efficiency, we first optimize nanopillar morphology for PL5 and PL6 color centers in SiC using FDTD simulations and investigate how color-center precision in nanopillars affects fluorescence intensity.

As shown in Fig.\ref{Figure1}(c and d), the collection efficiency depends on the top radius and height of the nanopillar structure. The white box in Fig.\ref{Figure1}(d) marks the optimal geometric morphology, which yields a collection efficiency of approximately 0.54 for PL6 color centers (0.32 for PL5 color centers, see Fig.S2(a) in Supplementary Material). This efficiency, however, decreases rapidly with imperfect alignment, i.e., when the color centers deviate from the optimal central position. Specifically, the doping region needs to be below 100 nm to approach optimal and uniform collection efficiency. Fig.\ref{Figure1}(f) clearly confirms this point by displaying the simulated low collection efficiency of bulk SiC crystal, random performance in non-aligned nanopillars, and the optimal and uniform performance in aligned nanopillars with a 90 nm doping region.

We use a self-aligned fabrication process based on bilayer photoresist to achieve optimal size control of nanopillars and precise alignment of color centers with the nanopillars (Fig.\ref{Figure2}(a)), thereby  obtaining the best fluorescence collection efficiency. We first define a hole array pattern with a pitch of 90 nm in the top photoresist layer via electron beam lithography. This pattern serves two purposes: it acts as a doping mask for ion implantation and as an etch mask for wet etching of the bottom photoresist layer. By controlling the isotropic wet etching of the bottom layer, the holes expand laterally to define the dimensions of the metal mask for subsequent nanopillar etching, enabling precise control of the nanopillar diameter. With the two photoresist patterns confining the ion implantation region and the etching region separately, this process ultimately achieves scalable positioning of the color-center-doped region at the center of each nanopillar.

Fig.\ref{Figure2}(b) shows an SEM image of a typical nanopillar before the mask removal and surface cleaning. By precisely controlling the hole size during wet etching and the etching depth of SiC, the final nanopillar has a top radius of $\sim$310 nm and a height of $\sim$480 nm, matching the optimal geometric morphology identified in Fig.\ref{Figure1}(d). Fig.\ref{Figure2}(c) presents the uniform morphology of the nanopillar array, confirming the scalability of this fabrication method. The PL5 and PL6 color centers are then formed by the subsequent 15 keV/3 keV oxygen ion implantation and high-temperature annealing.

\subsection{Statistical Characterization of Fluorescence Intensity in Nanophotonic Arrays}

Fig.\ref{Figure3}(a) shows a confocal image of the nanophotonic array, in which typical PL5 and PL6 centers are marked by their distinct zero-field ODMR spectra and fluorescence polarization. Fig.\ref{Figure3}(b) presents the fluorescence saturation curves of these representative color centers, along with background count rates rigorously derived from the second-order correlation function $g^{(2)}(0)$ (see Supplementary Material for the evaluation method). The results show that PL5 and PL6 color centers achieve saturation fluorescence count rates of 1.25 Mcps and 2.20 Mcps, respectively, which agree well with the simulated collection efficiency.
The background fluorescence from the nanopillar structures is negligible. Fig.\ref{Figure3}(c) and (d) show the zero-field ODMR spectra and fluorescence polarization results of PL5 and PL6 color centers in the nanopillar structures. Their ODMR contrasts show no significant difference from previously reported values for PL5 and PL6 in bulk \textsuperscript{\cite{Li2022NSR,hu2026high,chen2026oxygen}}.

\begin{figure*}
\centering 
\includegraphics[width=0.9\textwidth]{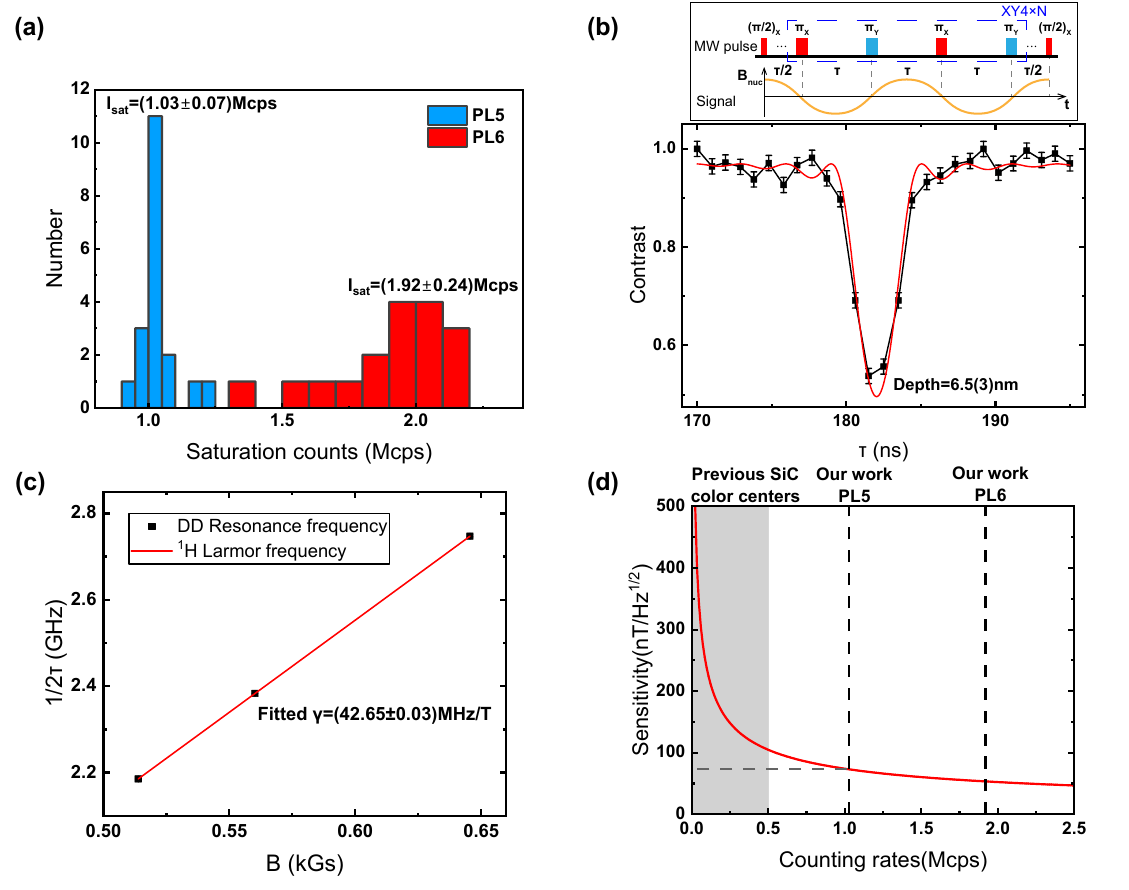}
\caption{Proton NMR signal detection and sensitivity improvement.
(a) Statistics on saturation count rate of PL5 (blue) and PL6 (red). Average saturation count rates and standard deviation for PL5 and PL6 are marked in the figure, respectively.
(b) Detection of proton nuclear signals in immersion oil using a PL5 center at a depth of 6.5(3) nm. Error bars represent standard errors of measured data. The depth of the PL5 center is determined using the previous NMR method\textsuperscript{\cite{pham2016nmr}}. The schematic shows the XY4-N dynamical decoupling (DD) sequence used in nuclear signal detection.
(c) Calibration of the Nuclear-Spin Gyromagnetic Ratio. The resonance frequency exhibits an excellent linear dependence on the axial magnetic field, and the extracted gyromagnetic ratio is consistent with that of the $^1 H$ nuclear spin.
(d) The sensitivity improvement of PL5/PL6 in nanophotonic structures compared to previous work.
 }
\label{Figure4}
\end{figure*}

Fig.\ref{Figure4}(a) displays the statistical analysis results of fluorescence intensity. 
We randomly selected 19 PL5 and 17 PL6 color centers in total. The statistics confirm that the saturation fluorescence count rates of both color centers are highly uniform: the average fluorescence count rate is 1.03 Mcps for PL5 color centers with a standard deviation of 0.07 Mcps, and 1.92 Mcps for PL6 color centers with a standard deviation of 0.24 Mcps. Compared with typical photon count rates on the order of 100 kcps for PL5 and PL6 color centers reported in previous studies \textsuperscript{\cite{wornle2026scalable,Li2022NSR,chen2026oxygen,hu2026high}}, the count rates of PL5 and PL6 color centers in the quantum devices fabricated in this work can be stably increased to 1-2 Mcps.

\subsection{Nanoscale NMR Detection}

The measured high ODMR contrast, bright fluorescence intensity, and the shallow depth of color centers establish these devices as sensitive nanoscale quantum sensors. To verify this capability, we employ near-surface PL5 centers for the nanoscale detection of nuclear magnetic resonance (NMR) signals from the immersion oil introduced onto the nanopillar surface.

As shown in Fig.\ref{Figure4}(b), we use an XY4-30 dynamical decoupling (DD) sequence to detect the proton signal in a nanoscale volume, and evaluate the sensor’s distance from the surface via the known proton density and signal intensity\textsuperscript{\cite{pham2016nmr}}. Under a magnetic field of 645.58 G, a proton NMR signal at the corresponding frequency of $f=\frac{1}{2\tau}= 2.75 MHz$ is observed. This is further confirmed by determining the gyromagnetic ratio $\gamma = 42.65 MHz/T$ of the proton via measurements at different magnetic fields (Fig.\ref{Figure4}(c)). By fitting the proton signal, we determine a depth of $6.5(3) $ nm for the PL5 center beneath the device surface, which is consistent with the 3 keV ion implantation.

Benefiting from the strong fluorescence enhancement, the NMR spectra in Fig.~\ref{Figure4}(b) and (c) achieve a high signal-to-noise ratio (SNR). To evaluate the resulting gain in near-surface quantum sensing sensitivity, we compare the magnetic field sensitivity in Fig.~\ref{Figure4}(d), whereas the saturation fluorescence count rate was $\leq 500~\mathrm{kcps}$ and mostly $\sim 100~\mathrm{kcps}$ in previous work on SiC color centers\textsuperscript{\cite{widmann2015coherent,radulaski2017scalable,Li2022NSR,wornle2026scalable,chen2026oxygen,hu2026high,zhou2023plasmonic,sardi2020scalable}}. As the sensitivity scales as $\mathrm{\delta B \propto \frac{1}{\sqrt{I}}}$ where I is the photon counting rate, we achieve a 2-3 fold improvement compared to previous work in PL5/PL6 NMR detection\textsuperscript{\cite{chen2025single}}. The sensitivity reaches below $100nT/Hz^{1/2}$ for a representative PL5 color center, with $T_2$ reaches $20.2 \mu s$ and Rabi contrast reaches $26.7\%$ (see Supplementary Material for details). By further using shallower color centers via low-energy ion implantation, the SiC quantum devices demonstrated here establish a promising route to scalable nanoscale quantum sensing.

\section{Discussion}

This work addresses two long-standing bottlenecks of SiC color-center sensing, including weak fluorescence collection and random emitter positioning, through a self-aligned fabrication strategy that places shallow PL5 and PL6 centers precisely at the centers of optically optimized 4H-SiC nanopillars. The resulting arrays are scalable, ultra-bright (∼1-2 Mcps), and highly uniform across devices, with background fluorescence from the nanostructures being negligible. The precise alignment enables efficient parallel nanoscale NMR measurement under almost the same fluorescence intensity. Compared with previous SiC color-center work, the devices deliver a 2-3 fold improvement in magnetic field sensitivity, reaching below $100nT/Hz^{1/2}$, a level comparable to the state-of-the-art nanoscale quantum sensing using NV centers in diamond \textsuperscript{\cite{maletinsky2012robust,rovny2024nanoscale}}. The combination of high brightness, uniformity, and scalable fabrication positions these nanopillar arrays as practical building blocks for AFM-based quantum probes and large-scale nanoscale sensing.

\section{Materials and Methods}

\subsection{Simulation}
We use the finite-difference time-domain (FDTD) method to simulate the enhanced photon collection efficiency of the nanopillar and design its geometry, in which the top radius and height of the nanopillar are variables. The sidewall angle is a fixed value of $78^{\circ}$, influenced by etching conditions.

The collection efficiency is expressed as $\eta $ = $I$/$I_0$ , where $I$ is the intensity integral of the far-field electric field within the lens collection angle, and $I_0$ is the source power. The position of the color center is set as $x=0, z=10 nm$, representing a precise implantation at the nanopillar center of a shallow PL5/PL6. The wavelength of the electric dipole is set to 1040 nm, close to both zero-phonon lines of PL5 (1043 nm) and PL6 (1038 nm). The refractive index of material is set as $n_x = n_y = 2.58, n_z= 2.63$, represents $n_o$ and $n_e$ of 4H-SiC at this wavelength, respectively.

\subsection{Device Fabrication}
The sample used in the experiment is from a 4H-SiC wafer with a silicon surface intrinsic epitaxial layer of 8 $\mu m$ thickness from Xiamen Powerway Advanced Material Co., Ltd., which is diced and polished to 30 $\mu m$ thickness by chemical mechanical polishing (CMP). Before the experiment, the sample is cleaned in a \SI{180}{\celsius} triacid (perchloric, nitric, and sulfuric acid) bath.

A PMGI-SF6 layer of ~280 nm thickness and a PMMA-A4 layer of ~200 nm thickness are spin-coated on the 4H-SiC surface, along with the conductive polymer layer as an anti-charging layer. Electron beam lithography (EBL) and development in MIBK: IPA are performed to make 90 nm hole array patterns in the PMMA layer. Isotropic wet etching in AZ300MIF is then performed to the PMGI layer through holes in PMMA. Implantation of oxygen ions with an energy of 15 keV and a dose of $2 \times 10^{11}cm^{-2}$, and energy of 3 keV and a dose of $10^{12}cm^{-2}$, is applied on 2 different regions of the sample. The PMMA layer is then removed in acetone and cleaned in isopropanol (IPA). A chromium layer of $\sim$60 nm is deposited on the surface by electron beam evaporation, and the lift-off process is performed in N-methyl pyrrolidone (NMP). Reactive ion etching with inductively coupled plasma(ICP-RIE) is carried out with $\mathrm{S F_6:O_2}$, 40:2 standard cubic centimeters per minute (sccm) mixed gas to form the nanopillars. Triacid, piranha, and HF are used to remove the chromium mask and reduce the fluorescent background. Finally, PL5/PL6 are formed by annealing at \SI{1050}{\celsius} in vacuum.

\section{Acknowledgments}
This work is supported by the National Natural Science Foundation of China (Grants No. T2325023, 92265204, 92565304, 12474500), Quantum Science and Technology-National Science and Technology Major Project (Grant No. 2021ZD0302200), and the National Key R\& D Program of China (Grants No. 2023YFF0718400). The fabrication process was partially performed at the USTC Center for Micro and Nanoscale Research and Fabrication. 

Author contributions: Y.W. and M.W. proposed the idea and supervised the experiment. Xin.C. and H.S. performed substrate preparation. M.W. and Chenyang.L. designed the fabrication process. Chenyang.L. performed the simulation. Chenyang.L., Cunliang.X. and H.Z. performed the fabrication process. Chihang.L. collected and analyzed the experimental data. Chihang.L. and F.S. constructed the experimental setup. Y.W., M.W., Chenyang.L. and Chihang.L. wrote the paper. All authors discussed the results and commented on the manuscript.

\section{Conflict of interest}
The authors declare that they have no conflicts of interest.

\bibliography{main}
\end{document}